# Foliar area measurement by a new technique that utilizes the conservative nature of fresh leaf surface density.

*O.S. Castillo[1], E.M. Zaragoza[1], C. J. Alvarado[1], M. G. Barrera[1] and N. Dasgupta-Schubert[1, 2#]* *

1. Institute of Chemical Biology (IIQB), University of Michoacán (UMSNH), Cd. Universitaria, Morelia, Mich. 58060, México.

2. Dept. Environmental Biotechnology, Helmholtz Centre for Environmental Research (UFZ), Permoserstraße 15, 04318 Leipzig, Germany

*([#] On sabbatical leave from UMSNH)*

* Author for correspondence: nita@ifm.umich.mx; nabanita.dasgupta-schubert@ufz.de

Tel/Fax: +52-443-326-5788/5790 (Mexico); +49- 0341 235 ext. 1765 (Germany)

Co-authors:

Omar Surisadai Castillo Baltazar: omarscb@gmail.com

Esther Magdalena Zaragoza Arias: ibq.esther@inbox.com

Carlos Juan Alvarado López: charliealvarado@gmail.com

Maria Guadalupe Barrera Arguello: barreraloop@gmail.com

**Abstract**

Leaf-area (LA) is a plant-biometric index important to agro-forestry and crop production. Previous works have demonstrated the conservativeness of the inverse of the product of the fresh leaf's density and thickness - the so-called 'Hughes' constant' ($K$). We use this fact to develop 'LAMM', an *absolute* method of LA measurement (*i.e.* no regression fits or prior calibrations with planimeters). Nor does it require drying the leaves. The concept involves the *in situ* determination of $K$ using geometrical shapes and their weights obtained from a sub-set of fresh leaves of the set whose areas are desired. Subsequently the LAs (at any desired stratification level), are derived by utilizing K and the previously measured masses of the fresh leaves. The concept was first tested in the simulated ideal case of complete planarity and uniform thickness by using plastic-film covered card-paper sheets. Next the species-specific conservativeness of $K$ over individual leaf zones and different leaf types from leaves of plants from two species, *Mandevilla splendens* and *Spathiphyllum wallisii*, was quantitatively validated. Using the global average $K$ values, the LA of these and additional plants, were obtained. LAMM was found to be a rapid, simple, economic technique with accuracies, as measured for the geometrical shapes, that were comparable to those obtained by the planimetric method that utilizes digital image analysis (DIA). For the leaves themselves, there were no statistically significant differences between the LAs measured by LAMM and by the DIA and the linear correlation between the two methods was good ($R^2 = 0.99$).

## 1. Introduction

Leaves are of fundamental importance to plants. They constitute the plant's power generation and aerial environmental sensing units. The amount of photosynthetic light harvested depends directly on the leaf-area (LA), which affects plant growth and bio-productivity and hence also the agro-economic return from the crop. Additionally, leaf biometrics are markers of nutritional status or environmental stress on the plant (Meziane & Shipley, 1999; Vile *et al.*, 2005). The leaf area index (LAI) of a stand of plants is one of the most frequently used parameters for the analysis of canopy structure and is an important structural characteristic of crop/forest monitoring and crop productivity (Behera et al, 2010). It is fundamentally important as a parameter in land-surface processes and for parametrizations in climate models. Direct methods of LAI determination are based on the destructive technique of the LA measurements of the individual leaves and are considered to be the most accurate, albeit time-consuming, ways of determining LAI (Jonckheere et al, 2004). Nonetheless, they are used to benchmark non-destructive methods of LAI determination (Behera et al, 2010; Kirk et al, 2009).

Hence the measurement of LA occupies an important place in the slew of plant biometric techniques. This importance has stimulated the proposition of a variety of methods for LA measurement (Coombs et *al*., 1985; Mohsenin, 1986; Sharatt & Baker, 1986; Ma *et al*., 1992; Singh *et al*., 1995; Korva & Forbes, 1997; O'Neal *et al*., 2002; Igathinathane *et al*., 2006; Rico-Garcia *et al*., 2009; for a review of LAI determination, see Jonckheere *et al*., 2004). The method that has emerged as the most frequently used, particularly for herbaceous broad-leaved species, and which has been commercialized, is planimetry (Jonckheere et al, 2004): the leaf, excised at its junction to the petiole, is laid on a scanner-type bed, and the pixel count of its digital image using appropriate software, is used to quantify its area (*e.g.* O'Neal *et al*., 2002 and references therein; LI-3000®, Li-Corr, Lincoln, NE, USA). While this method - henceforth called the DIA - is well-established, some difficulties persist. Leaves with non-planar laminae (wavy leaves) do not sit flush against the bed so that the kinks could cause zones to be occulted; the pixel contrast for leaves with insufficient chlorophyll content (yellowish leaves) is sometimes not picked up accurately enough by the software. Both can result in an underestimation of the area (Tsuda, 1999; Rico-Garcia *et al*, 2009). The problems can be corrected to some extent by positioning and flattening the leaves manually and by manual grey-scale corrections of the digitized image, but both detract from the speed of the procedure. The need to find a sufficiently accurate non-DIA based method of determining leaf areas using functional traits of leaves but which does not rely on prior statistically fitted 'calibrations', *i.e.* an absolute as against a relative method, served as the central motivation for this work.

In the context of a relative determination, it is germane to briefly recapitulate the second traditionally used method of LA determination - the gravimetric method (Jonckheere et al, 2004). The method consists of determining the fresh LAs of a sub-sample of leaves of a particular species drawn from the global field sample using a DIA method, correlating these with their dry masses and then applying the LA to dry leaf mass ratios or specific leaf area (SLA), to the dry masses of all the harvested leaves to obtain their areas. The method is not stand-alone (i.e. it requires a prior instrument of LA determination), it involves extra processing (oven-drying) and the SLA is not a well-conserved quantity (see below) in that it has been observed to possess wide spatial and temporal variations in many tree species (Jonkheere et al, 2004; Fila and Sartorato, 2011). However it is a convenient (more rapid) method for the estimation of LAI for very large leaf samples (Jonckheere et al, 2004).

In an article published in 2002, Roderick & Cochrane tested and corroborated the observation made earlier by Hughes *et al.* (1970). Their combined measurements covered a large number of plant species. They observed that for herbaceous species there apparently exists a unique species-specific relationship between the area and the mass of *fresh* leaves.

The relationship connecting the leaf area (projected normal to the surface), $A$ (m$^2$) and the fresh leaf mass $M$ (kg) is,

$$A = KM \quad (1)$$

Where $K$, termed the 'Hughes' Constant' by Roderick & Cochrane (2002), is related to the density ($\rho$) and thickness ($\tau$) of the fresh leaf as,

$$K = \frac{1}{\rho \tau} \quad (2)$$

Since the dimension of $K$ is [square of length·mass$^{-1}$], it can be construed as the fresh leaf specific area, or in analogy with thick-film materials technology, the inverse of the fresh leaf surface density.

Using measurements of the LA and leaf mass from 15 broad-leaved species Roderick and Cochrane (2002) showed that for each species $K$ was approximately a constant despite variations in leaf thickness and water content. The latter would significantly affect $\rho$. Based on these results, they contend that for a given species, the Hughes' constant is likely to be much more conservative than other functional attributes, *e.g.* leaf area per unit dry mass (the so-called specific leaf area, SLA), leaf water content, *etc*. This is a remarkable finding. It implies that despite factors affecting $\rho$ and $\tau$ separately, the plant dynamically adjusts both through their functional connectedness, to maintain $K$ as constant.

The utility of conserved quantities is that they allow accurate measurements to be made. The analogy to the leaf in materials science is the thick self-supporting film whose density is constant and known. The thickness of the film can then be determined accurately by punching/cutting out a geometrical shape of known area, weighing it and thereupon utilizing equations 1 and 2. Using the property of the conservativeness of $K$ and an approach inspired by thick-film technology, we now proceed to describe the development of the LAMM method of LA measurement (Leaf Area Measurement by Mass). The work is presented in the five stages of (i) proof of concept using thin plastic-film covered card-paper sheets of uniform thickness; (ii) the checks on the conservativeness of $K$ over differently sampled areas of the given leaf and (iii) for different leaf types; (iv) the estimation of the accuracy of the method; and (v) the determination of the areas of leaves of a different individual of the same species using the $K$ from a different plant, which tests the constancy of $K$ for the species and its exportability from one plant to another. Plants of two different species, *Mandevilla splendens*, common name Pink-allamanda and *Spathiphyllum wallisii*, common name Peace-lily, were used. At every stage, measurements by DIA on the same specimens were made to benchmark the LAMM against this standard method.

## 2.   Conceptual description

Let us suppose that there are $v$ broad-leaved plants of a particular species in an experimental set, whose total foliar area (FA) is desired to be measured. The index $j$ for the individual plants runs from 1 to $v$ ($1 \leq j \leq v$). Further, each plant has $n$ leaves. The Hughes constant for the $i^{th}$ leaf of any plant where $1 \leq i \leq n$ is,

$$K_i = \frac{1}{\rho_i \tau_i} \tag{3}$$

The $i^{th}$ leaf is detached at its base from the petiole and weighed, to yield the mass $M_i$. Now, using a sharp-edged die-punch of known ID, discs are punched out in a random manner over the face of the leaf, taking care to see that ribbed and fleshy areas are equally sampled. In the general case, several die-punches of differing ID may be used. The punch-out process will result in a maximum of $P$ discs where the ID of the $p^{th}$ disc ($1 \leq p \leq P$) is $d_p$ and its mass as measured in the weighing balance is $m_p$. Its area $a_p$ then is,

$$a_p = \pi \left(\frac{d_p}{2}\right)^2 = K_i m_p \tag{4}$$

which can be easily calculated.

If all the $P$ discs are pooled together and weighed to yield the mass $m$,

$$m = \sum_{p=1}^{p=P} m_p = \left(\frac{1}{K_i}\right) \sum_{p=1}^{p=P} a_p \tag{5}$$

or,

$$K_i = \left(\frac{1}{m}\right) \sum_{p=1}^{p=P} a_p \tag{5}$$

Eqns. 4 and 5 assume that the Hughes constant is invariant over the face of the leaf and has the single value $K_i$, making it applicable for all the discs. This assumption is tested in section 3 and found to be true, in section 4.

The area of the $i^{th}$ leaf is then,

$$A_i = K_i M_i \tag{6}$$

Doing likewise for all the n leaves of the plant we will obtain their different values of $K$ and their different areas. We will then obtain the total FA of the plant, $A$, by summing over all the different areas $A_i$,

$$A = \sum_{i=1}^{i=n} A_i = \sum_{i=1}^{i=n} K_i M_i \tag{7}$$

Roderick & Cochrane (2002) have pointed out that the Hughes constant is nearly invariant with respect to leaf type in the given plant – an observation that we corroborate (sections 3 and 4).

Hence,

$$K_1 = K_2 = ... = K_i = ... = K_n = K \tag{8}$$

Thus a fine-tuned 'stratification' of the value of the Hughes constant leaf by leaf, becomes unnecessary and the procedure can be shortened. One would only need to sample a much smaller number of arbitrarily selected leaves of the plant, punch out the discs from these as stated above, weigh them and obtain the value of $K$ which is now the Hughes constant representing the whole plant. We call it the global average $K$ ($K_G$). Alternatively, a value for $K_G$ may be derived as the average of the measured values of $K$ corresponding to arbitrarily selected sub-sets of different leaf types (young, old, *etc*.).

Then,

$$A = \sum_{i=1}^{i=n} A_i = \sum_{i=1}^{i=n} K_i M_i = \sum_{i=1}^{i=n} KM_i = K \sum_{i=1}^{i=n} M_i = KM = K_G M \tag{9}$$

where $M$ is the weight of all the detached and intact leaves of the plant, obtained by weighing prior to the punch-out of discs from some of them.

For the laboratory experiment that consists of the $v$ number of plants, the area $A$ in eqn. 9 is actually the FA of the $j^{th}$ plant of the set, $A_j$ that has the total leaf mass $M_j$ and the Hughes constant $K_{Gj}$. The FA of all the $v$ plants of the set taken as a single entity, $A$, is then the sum over all $A_j$

$$A = \sum_{j=1}^{j=v} A_j = \sum_{j=1}^{j=v} K_{Gj} M_j \tag{10}$$

Roderick & Cochrane (2002) observe that the Hughes constant remains approximately the same for all plants of the same species (but varies between species). In that case, 'stratification' on a plant by plant basis to obtain the value of $K_{Gj}$ for each is not required and the procedure can be further simplified. The overall global value of $K$ for all the plants in the set, $K_G$, can be obtained by punching out discs from randomly selected leaves of different plants in the set using die-punches of known IDs in the manner described before. Prior to the punch-out operation, all the detached leaves of all $v$ plants are pooled and weighed to yield the mass $M$. Therefore, the total FA of all $v$ plants becomes simply,

$$A = \sum_{j=1}^{j=v} K_{Gj} M_j = K_G \sum_{j=1}^{j=v} M_j = K_G M \tag{11}$$

The process is simple and rapid. In contrast, for a typical DIA measurement, the detached leaves have to be placed carefully on the scanner-bed which itself might be space limited to only a certain number of leaves at a time and furthermore, the measurement result depends on the analysis of the image of each leaf. This is the likely reason why for large sample sizes, the DIA is not considered to be the most efficient LA measurement technique and the gravimetric method, in spite of its deficiencies (see section 1, p. 4), is considered as a viable option (Jonckheere et al, 2004).

## 3. Materials and Methods

### 3.1 Proof of Concept

Dark and light green coloured standard card-paper sheets (Pacon Corporation, USA) were tightly covered with plastic film (Parafilm® , Pechiney Plastic Packaging Inc., USA). They served as simulators of two levels of the chlorophyll of leaves together with their waxy cuticles. Stainless steel die-punches of IDs 1.3, 1.0 and 0.5 cm (Leon Weill S.A., Mexico) were used to punch out several discs from these sheets. Each set of discs of a particular diameter and particular shade of green was divided into the sub-sets I and II. The masses of the all the discs in each subset were obtained by weighing the discs on a laboratory balance (Ohaus Adventurer™ SL, Ohaus Corp., USA). The subset I was used to determine the $K$ value (*cf*. section 2) of the sheet that it was punched out from. Thus $K$ values for the dark and light green sheets were obtained as the values distinguished by the diameters of the disc. The discs of sub-set II were designated as the model "leaves" of unknown areas. For the sheet of a given shade of green, the $K$ value for a particular disc diameter was now used to calculate the "leaf areas" of the discs of set II with non-identical disc diameters. The idea behind using the $K$ values obtained from discs of one size to determine the "leaf areas" of discs of another size, was to observe the variability in the area values consequent upon the possible variation of $K$ with respect to the radial size of the sampled zones over the face of the sheet (or the leaf in the real case as will se seen in the next section). Finally, the global average $K$ value ($K_G$) was computed as the average over all disc sizes for the particular sheet and this value was also used to obtain the "leaf areas" of the discs of the sub-set II.

The "leaf areas" of all the sub-set II discs were also obtained by using the DIA method as specified by O'Neal *et al*. (2002) that had been cross-corroborated by the use of the LI-3000® commercial leaf area meter. The discs were placed on the bed of the scanner (Canon PIXMA MP520®) under a blue paper background that was graduated along the X-Y axes to the maximum length of 15 cms. The colour photographs of 300 ppi resolution were exported to the desktop of a standard PC in the imaging format JPEG as required by the image analysis software. The latter was the software SIDELOOK® (Zehm *et al*., 2003; Nobis & Hunziker, 2005) that was used in the manner described in the software manual (http://www.appleco.ch/sidelook_sample.zip) to analyse the images and obtain their areas. In the case of the real leaves the software Image-J® was also used where the protocol as stated by its on-line manual (http://rsbweb.nih.gov/ij/docs/menus/analyze.html#ap) was followed to obtain the areas.

To calibrate the time factor involved in each type of measurement, the duration of measurement for all the light and dark green plastic covered discs of set II using the LAMM and the DIA (SIDELOOK) methods were timed.

The area results are shown in figures 1 and 2.

*3.2    The determination of K for plant leaves, the estimation of the accuracy of LAMM and the comparison of leaf areas obtained by LAMM and DIA*

Two nursery grown plants, one a four month old *M. splendens* and the other a two year old *S. wallisii* were selected for experimentation. The leaves selected were 6 of *M. splendens* that included the morphological types of 1 mature-chlorotic leaf, 2 mature and 3 young leaves, and 5 of *S. wallissi* that were all mature. Using the set of previously mentioned die-punches that now also included one of 0.3 cm ID, discs were punched out of these leaves in the manner explained earlier. The $K$ values for all the selected leaves of each plant were separately obtained for discs corresponding to each die-punch ID. The $K$ averages for each diameter for each leaf (morphological) type as well as the averages over all disc diameters for each leaf type were then computed. Finally the $K_G$ value was computed as the grand average over all disc diameters for all selected leaves of the given plant. The results are shown in table 1.

Prior to punching out the leaf discs, the areas of the 6 and 5 leaves respectively of *M. splendens* and *S. wallisii* were measured by the DIA using SIDELOOK® as well as Image-J®. Later, the areas of the punched out discs were also measured using the same softwares.

For each plant, the $K_G$ values were used to obtain the areas of the discs of the different diameters, averaged over all leaves. Only in the case of *M. splendens* the areas for discs from the mature-chlorotic leaf was excluded because Image-J produced an aberrant result. Since the geometrical area for each diameter is known, the accuracy of the LAMM and the DIA methods were obtained by comparing the absolute geometrical areas of the discs with the average LAMM derived and the average DIA derived areas. The results are shown in figures 3a and 4a for the two plants.

Now, utlizing the $K_G$ values, the area of each of the selected leaves of the two plants (including the mature-chlorotic one of *M. splendens*) was obtained by LAMM and compared to the areas obtained by the DIA methods. The results are shown in figures 3b and 4b.

*3.3 The determination of the areas of leaves of different plants of the same species using LAMM and DIA and their inter-comparison*

From two different individuals of the same plant species, *M. splendens* and *S. wallisii*, 9 leaves of the former and 5 of the latter were randomly selected. One of the 9 leaves of *M. splendens* was chlorotic. Using the $K_G$ values obtained from the same plant types of the previous set, the area of each of the currently selected leaves was obtained by the LAMM technique. Their areas had already been obtained by the DIA method. The results using each technique were compared. This is shown graphically in figures 5a and 5b.

For all measurements, the errors on primary determinations were computed as the standard errors, and for the determination of derived quantities, the standard formulae for error propagation were used (Mandel, 1984). Since the DIA measurement on each leaf as a single entity was a unique measurement, no statistical error was assigned to it. The expressions (12) and (13) were used to compute the % error (*E*) or relative deviation (*RD*) of the experimental quantity (*X*) with respect to the reference (*R*), and the % accuracy (*AC*) (Mandel, 1984).

$$E \text{ or } RD\,(\%) = \left|\frac{R-X}{R}\right| \times 100 \tag{12}$$

$$A\,(\%) = 100 - E\,(\%) \tag{13}$$

Furthermore, to determine whether the results between two sets of measurements (intra LAMM or inter LAMM and DIA) were significantly different statistically or not, the Student's t-test was applied (Mandel, 1984)

**4. Results and Discussion**

*4.1*   *Proof of Concept*

For the light green plastic film covered card-paper sheet, fig. 1 shows the areas of the discs of set II (the model "leaves") obtained by LAMM using the *K* values derived from measurements on set I using discs of different diameters. Also shown are the true geometrical areas of the set II discs, using which the % accuracies for the LAMM results have been derived. For this idealized case where the surface is homogeneous, planar and the thickness uniform, there is no clear dependence of the accuracy of the LAMM result on the dimension of the zone sampled to obtain the *K* value. The areas of discs obtained from *K* values derived from discs of small diameter (0.5 cm) give about the same accuracy as those derived using larger sized sampling zones. However, the global averaging over all sizes to yield the $K_G$ value, results in the best values for the accuracies (97-99%). The experiment using the plastic film covered dark-green card-paper shows the same results (not shown). However in this case the accuracies are shifted lower, to lie in the range of 85-97%. Possibly, a heavier application of the green dye to the paper to darken the hue, caused local inhomogeneities in the material's density to develop. Nonetheless, in this case also, the $K_G$ values resulted in the best range of accuracies (92-97%).

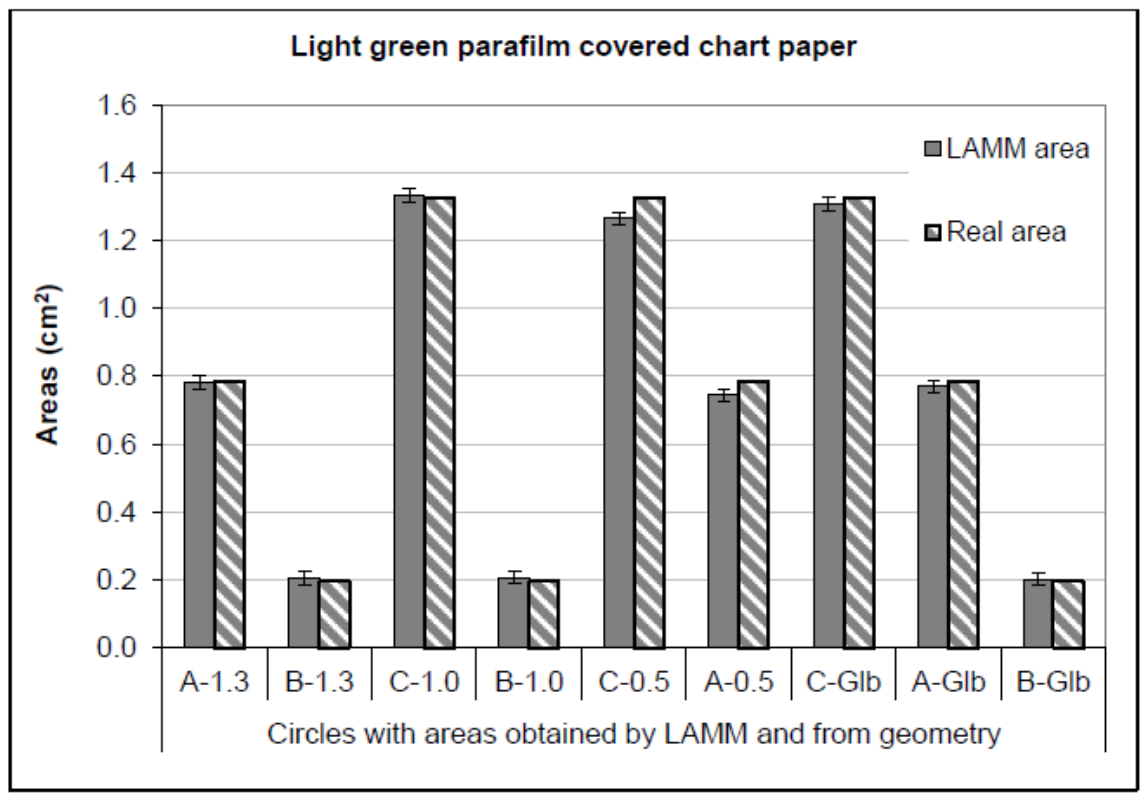

**Fig. 1**: A comparison of the areas of the discs of plastic-film covered light-green card-paper obtained by the LAMM

method with the actual geometrical areas. A, B and C are the punched discs of diameters 1.0, 0.5 and 1.3 cm respectively. The suffixes 1.3, 1.0, 0.5 and Glb refer to the fact of $K$ having been obtained from the different sized discs of diameters 1.3, 1.0 and 0.5 cms, while Glb refers to the global average $K$ value ($K_G$) obtained from all the discs. Proceeding from A-1.3 to B-Glb, the % accuracies for the LAMM are 99, 95, 99, 94, 95, 95, 99, 98 and 97.

The areas of the discs of the 3 different diameters 1.3-0.5 cm of set II obtained using the $K_G$ values from set I, were then compared to the DIA results of the areas using SIDELOOK®. The results for the light-green-plastic-film card-paper are shown in fig.2. The LAMM results are nearly identical to those for the DIA for this idealized case. The DIA infact produced accuracies in the range of 95-97% (fig. 2), slightly less than those of LAMM, and 94-98% for the equivalent case for the dark-green card-paper (not shown). The nearly identical results of the DIA for the two shades of green indicate that there are no spurious "white pixel" losses using SIDELOOK due to reflection or low chlorophyll content. The card-paper represents an idealized case for the DIA too, since the complete planarity presents a flat topography with no eclipsing errors in the image.

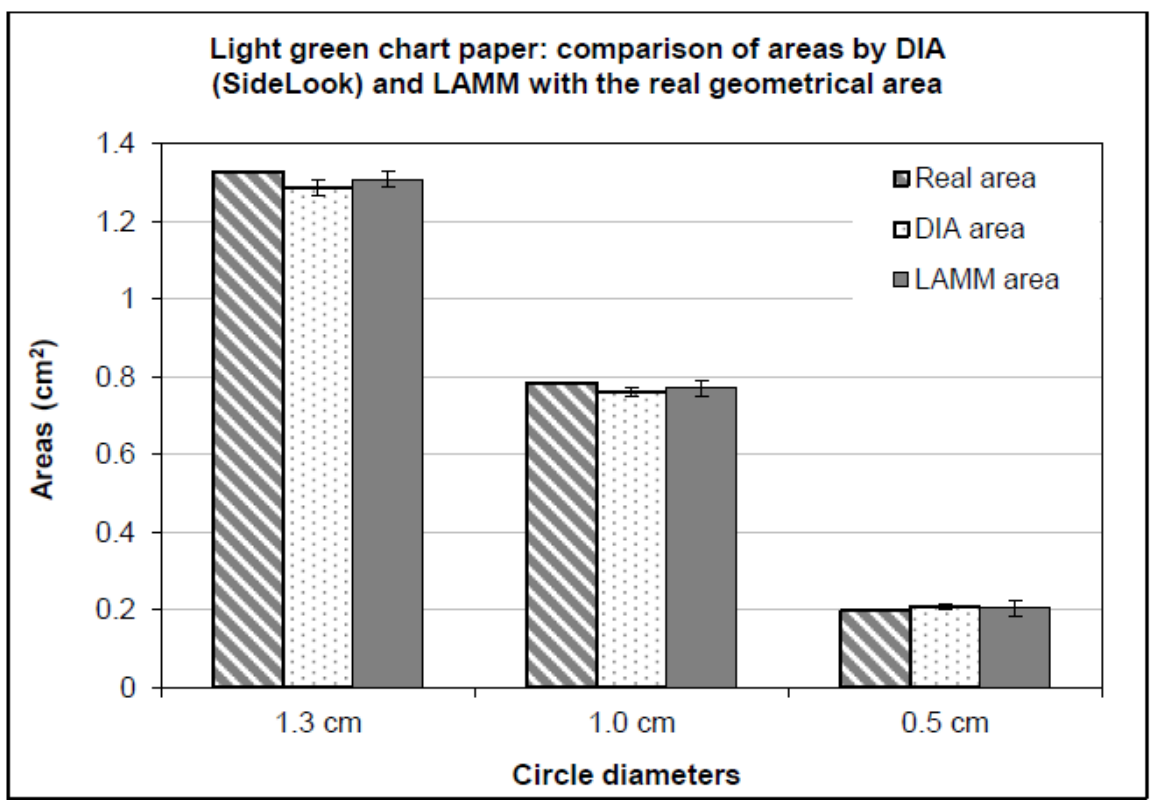

**Fig. 2**: The average areas of the plastic-film covered light green discs of diameters 1.3, 1.0 and 0.5 cm obtained

using the DIA method with SIDELOOK® (SL) and by the LAMM method using $K_G$. The DIA produced accuracies between 95 to 97% while the LAMM gave values between 97 and 99% (*cf.* figure 1).

Thus in the ideal situation, the LAMM technique proves to be at least as accurate as the standard technique of DIA. As far as the time efficiency of the two types of measurement is concerned, each SIDELOOK-DIA measurement took about 1 – 1.5 min to complete when no manual interventions for image corrections were utilized. The LAMM took on average about half this time. Hence the LAMM is also a rapid technique.

*4.2    The determination of K for plant leaves, the estimation of the accuracy of LAMM and the comparison of leaf areas obtained by LAMM and DIA*

The results for the determination of *K* from various sized discs obtained from leaves of different type from *M. splendens* and *S. wallisii* are summarized in table 1. The values for the inverse of *K* are shown because $1/K$ has the dimension of [mass·length$^{-2}$] which is the way the surface density of materials in thick film technology is described. Our first objective was to see whether in the case of real leaves the value of the Hughes constant varies across the face of the leaf *i.e.* with respect to the size of the zone sampled. The coefficient of variation (CV) is an indication of the random variability of the quantity under review, in other words its precision (Mandel, 1984). We see that the variability across the size of the zone sampled is not very different between the mature and mature-chlorotic leaves of *M. splendens* (8.2 and 10.4 %) but the variability is almost doubled for the young leaves (17.1 %) which suggests that for these developing leaves the Hughes constant is much more of a fluctuating quantity over the leaf face even while maintaining its average value close to those for other leaf types. The CVs for the global average surface densities for both plant species are close, in the environs of 11-12%. The CV for the global average surface density for the light-green card-paper sheet was 6.24%. The variability for real leaves is therefore only about double that of an industrially fabricated sheet where every attempt is made to keep the surface density constant. It shows the remarkable degree of regulation by plants in maintaining the uniformity of surface density (or its inverse, the Hughes constant) despite differences in leaf morphology, age, and state of health.

Our second and third objectives were to see if the Hughes constant varies with leaf type and if it varies between plant species. From table 1 we note that between the average values of 1/$K$ (indicated by ($K^{-1}$)$_{avg}$) for the three leaf types in *M. splendens* there exists no statistically significant difference. However, the global average value of 1/$K$ (($K^{-1}$)$_G$ in table 1) does differ statistically significantly between the two plant species. Both these results serve as quantitative corroboration of the observations made earlier by Roderick & Cochrane (2002) *viz*. that the Hugh's constant is robustly conserved within a species but it varies between species.

**Table 1** Values of the leaf surface density, $K^{-1}$ (kg·m$^{-2}$), obtained for the mature leaves with (MC) and without (M) chlorosis and young (Y) leaves of *M. splendens* and the mature leaves of *S. wallisii* with respect to the diameters, Dia. (cm), of the punched-out leaf discs used to obtain $K^{-1}$. The leaves are labeled by numbers, *L1*, *L2* etc. ($K^{-1}$)$_{avg}$ and ($K^{-1}$)$_G$ are respectively: the averages for each leaf-type over all the disc sizes, and the global average over all disc sizes and leaf-types for each plant species. CV(%) is the coefficient of variation for the various average values. The statistically non-significant, and significant, differences computed by the t-test, are indicated by * and ** respectively.

| | $K^{-1}$ x10$^2$ | | | | ($K^{-1}$)$_{avg}$ x10$^2$ | ($K^{-1}$)$_G$ x10$^2$ | CV (%) |
|---|---|---|---|---|---|---|---|
| | **Dia. (cm)** | | | | | | |
| | 1.3 | 1.0 | 0.5 | 0.3 | | | |
| | | | **Plant: *M. splendens*** | | | | |
| **Leaf type** | | | | | | | |
| *Mature-chlorotic (L1-MC)* | | | | | | | |
| | 3.16 | 2.55 | 3.06 | 2.67 | (2.86±0.30)* | | 10.4 |
| *Mature (L2-M, L3-M)* | | | | | | | |
| | 2.89±0.25 | 2.67±0.23 | 3.06±0.01 | 2.54±0.16 | (2.79±0.23)* | | 8.2 |
| *Young (L4-Y to L6-Y)* | | | | | | | |
| | 2.84±0.45 | 2.57±0.50 | 2.73±0.43 | 2.36±0.47 | (2.62±0.45)* | | 17.1 |
| | | | | | | (2.72±0.30)** | 11.2 |
| | | | **Plant: *S. wallisii*** | | | | |
| **Leaf type** | | | | | | | |
| *Mature (L1-M to L5-M)* | | | | | | | |
| | 2.15±0.23 | 2.14±0.43 | 2.10±0.28 | 1.72±0.19 | (2.03±0.25) | | |
| | | | | | | (2.03±0.25)** | 12.3 |

Figure 3a shows the average areas of the different sized discs obtained from the mature and young leaves of *M. splendens*, where for the LAMM determinations, the $K_G$ value had been used. The mature chlorotic leaf had been omitted because Image-J produced aberrant results for it. This was possibly because the areas with insufficient chlorophyll resulted in spurious "white pixels" that produced erroneous results for the area. This effect has also been noted by Tsuda (1999) when using an automatic leaf area meter that uses a DIA based technique. The accuracies for the two methods for all the discs, lay between 95-96% for the LAMM and 96-99% for the DIA. Figure 4a shows the same type of determinations for the mature leaves of *S. wallisii*. Again, the $K_G$ value had been used for the LAMM measurements. For these leaves the accuracies by the LAMM and by the DIA were 95-99% and 94-97% respectively. We note therefore, that for geometrical leaf areas obtained from both plant species, the LAMM and DIA accuracies are similar and show values that are close to those of the idealized case (section 4.1).

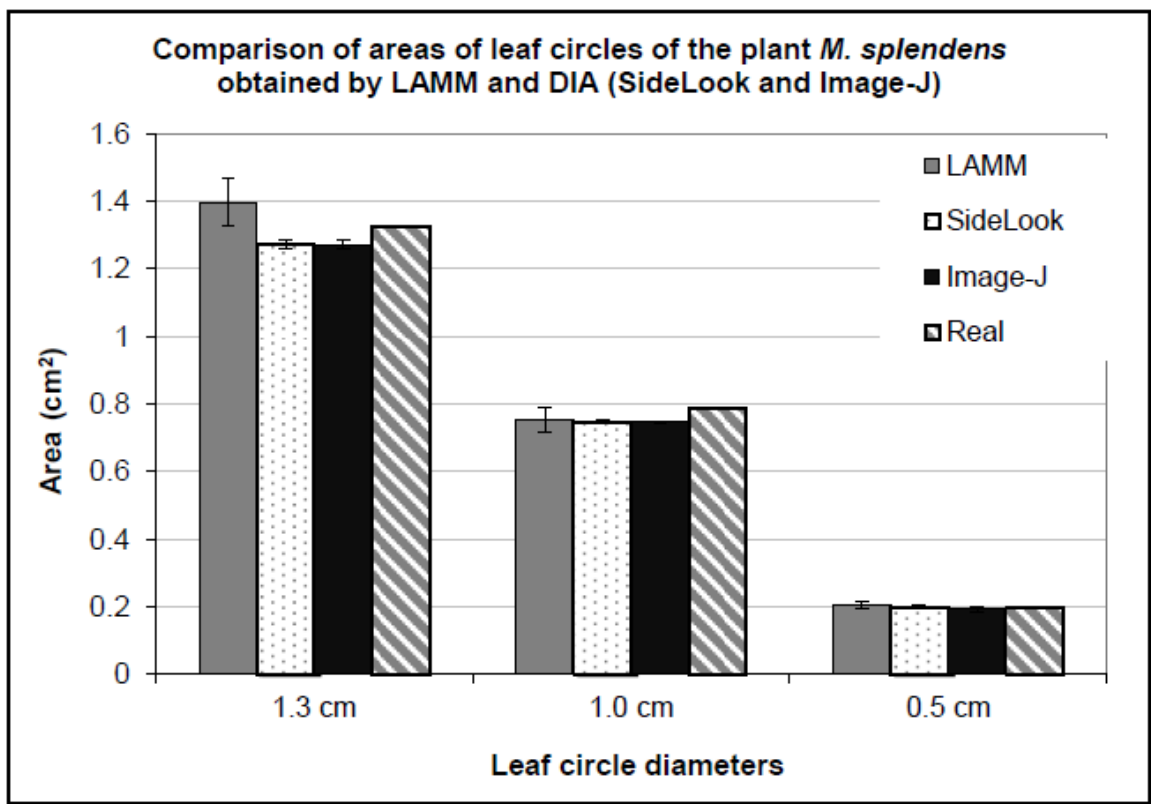

**Fig 3a** A comparison of the areas of leaf discs of diameters 1.3, 1.0 and 0.5 cm punched out from 2 mature (excluding the chlorotic) and 3 young leaves of *M. splendens* (*cf.* table 1) measured by the LAMM technique and the DIA using SL and Image-J® (IJ). The LAMM areas are the averages of all discs of the given diameter obtained using $K_G$. The DIA results pertain to the same discs as were used for the LAMM and are the averages over all discs of the given diameter. The accuracies of (a) the LAMM and (b) the DIA with SL and IJ, over the 3 disc sizes lay in the ranges of (a) 95-96% and (b) 96-99%.

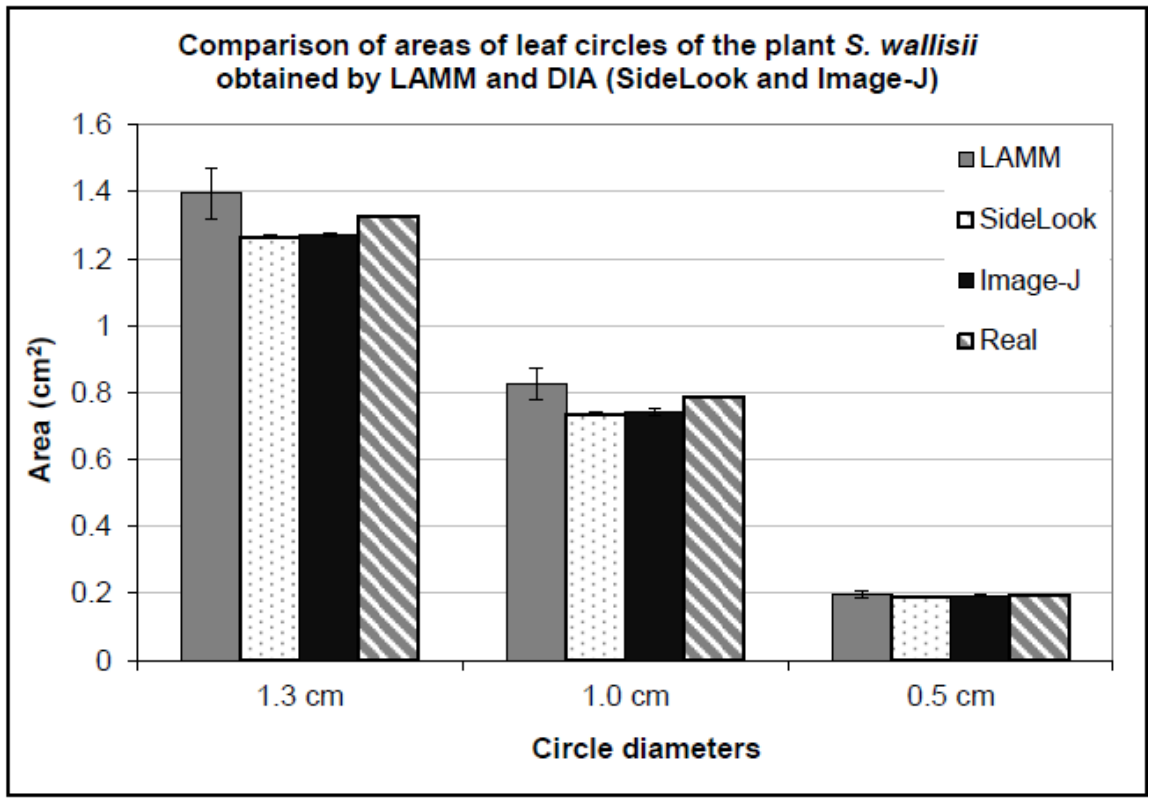

**Fig 4a** Same as fig. 3a but for the discs obtained from 5 mature leaves of *S. wallisii* (*cf.* table 1). The accuracies of (a) the LAMM and (b) the DIA methods using SL and IJ over the 3 disc sizes lay in the ranges of (a) 95-99% and (b) 94-97% respectively.

Figure 3b shows the areas of the 6 leaves of *M. splendens* (L1 to L6 with Y and M standing for 'young' and 'mature') obtained by LAMM using the $K_G$ value and by DIA. The mature-chlorotic (MC) leaf has also been included. The Image-J result for it clearly appears to be an underestimate. The average relative deviation between the LAMM and the SIDELOOK measured areas with the latter taken as the reference, was (10.4 ± 5.7)%. For real leaves of irregular shape the accuracies for the two methods cannot be quantitatively ascertained against the calculated geometrical area as was done for the leaf discs. One can only determine whether the area values are statistically close or not. The application of the Student's t-test to the overall average values of the areas obtained from the LAMM and the DIA indicated no statistically significant difference (95% confidence level (CL)) between the two.

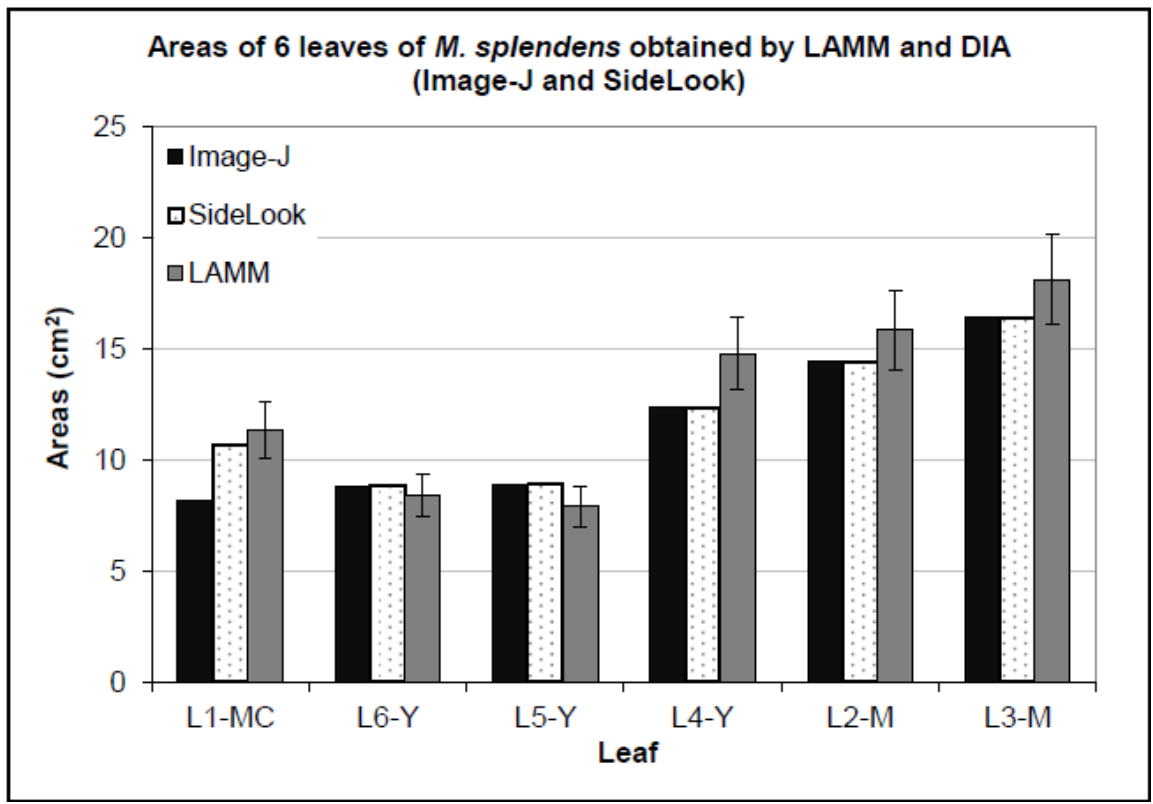

**Fig 3b**: A comparison of the areas of the 6 selected leaves, L1 to L6, of the plant *M. splendens* obtained by LAMM using the $K_G$ value (cf. table 1) and the DIA techniques using SL and IJ. The overall RD of the LAMM with respect to the SL is (10.37 ± 5.69) %. The Student's t-test indicated no statistically significant difference between the LAMM and DIA values of the overall average of the areas.

Figure 4b shows the analogous results for the 5 mature leaves of *S. wallisii*. For this plant the average relative deviation between the LAMM and SIDELOOK measured areas was (9.3 ± 4.9)%. As in the case of *M. splendens*, there was no statistically significant difference (95% CL) between the overall average of the leaf areas obtained by the LAMM and by the DIA.

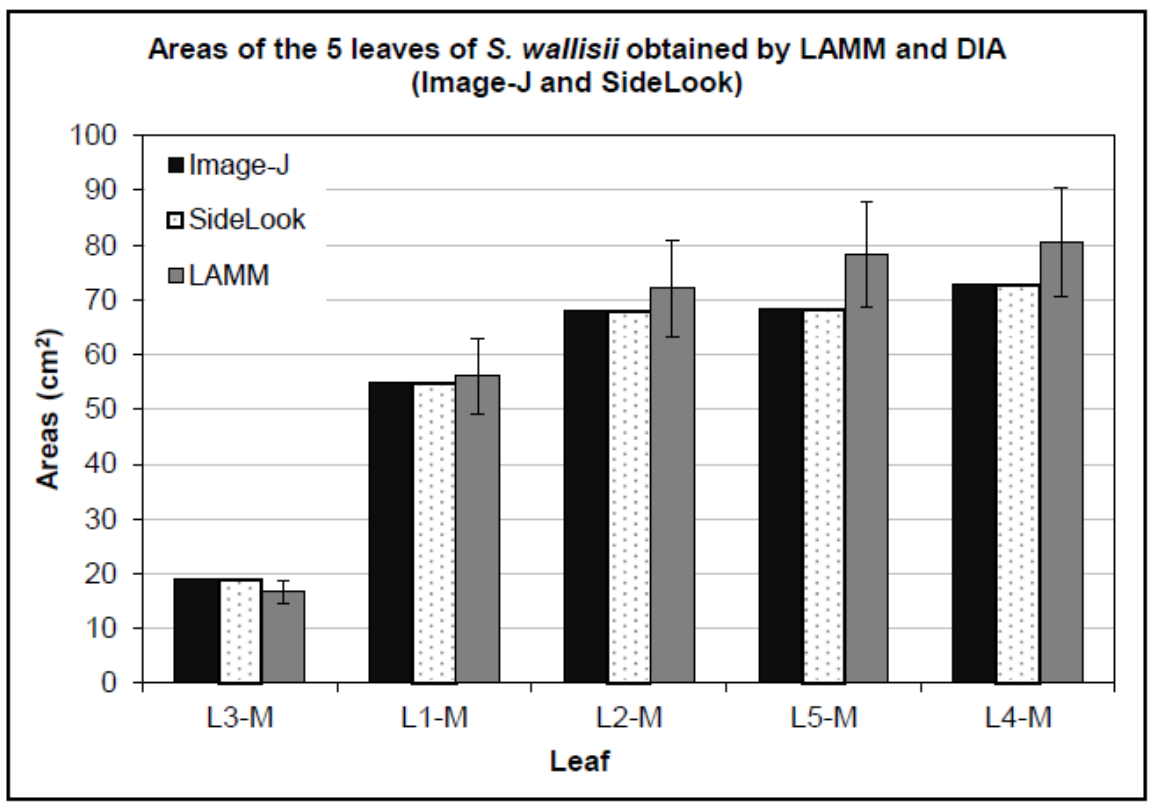

**Fig 4b** Same as figure 3b, but for the plant *S. wallisii* (*cf.* table 1). The overall RD of the LAMM with respect to the SL is (9.25 ± 4.90) %. As in fig. 3b, the differences in the overall average values of the areas obtained by the LAMM and by the DIA, were not statistically significant.

For both plants we notice that (i) the Image-J and the SIDELOOK (barring the case for the chlorotic leaf) produce identical results and that (ii) these in general are lower than the corresponding LAMM values. The first observation indicates that the algorithm used in both softwares perform the same operations, but since the version of SIDELOOK used here allowed a manual grey-scale correction, underestimations of the areas of leaves with chlorophyll deficiencies could be avoided. There are two possible reasons for the second observation. It was noticed that it was not possible to make all leaves lie flat on the scanner bed because of their rigidity and non-planarity. This would have caused some small portions in the ensuing images to be possibly "eclipsed" resulting in an underestimation of the areas by the DIA. The same cautions have been indicated by other researchers (Tsuda, 1999; Rico-Garcia *et al.*, 2009). On the other hand, while sampling leaf surfaces for the extraction of the Hughes constant value in the LAMM method, the tendency - albeit unintentional - is to under-include the ribbed zones where the leaf thickness $\tau$ is slightly higher. The lack of a thorough representation of these regions could mean that the $K_G$ value tends to shift slightly higher, resulting in a small increase of the area. An automation of the LAMM technique would help to improve the randomization of the sampling. In anycase, this effect is likely to be small and as seen for the case of the geometrical discs extracted from leaves (preceding paragraph), the intrinsic accuracy of the LAMM technique is high.

Thus for real leaves the LAMM method produces values not very different from the DIA and with accuracies of the same order of magnitude.

*4.3    The determination of the areas of leaves of different plants of the same species using LAMM and DIA and their inter-comparison*

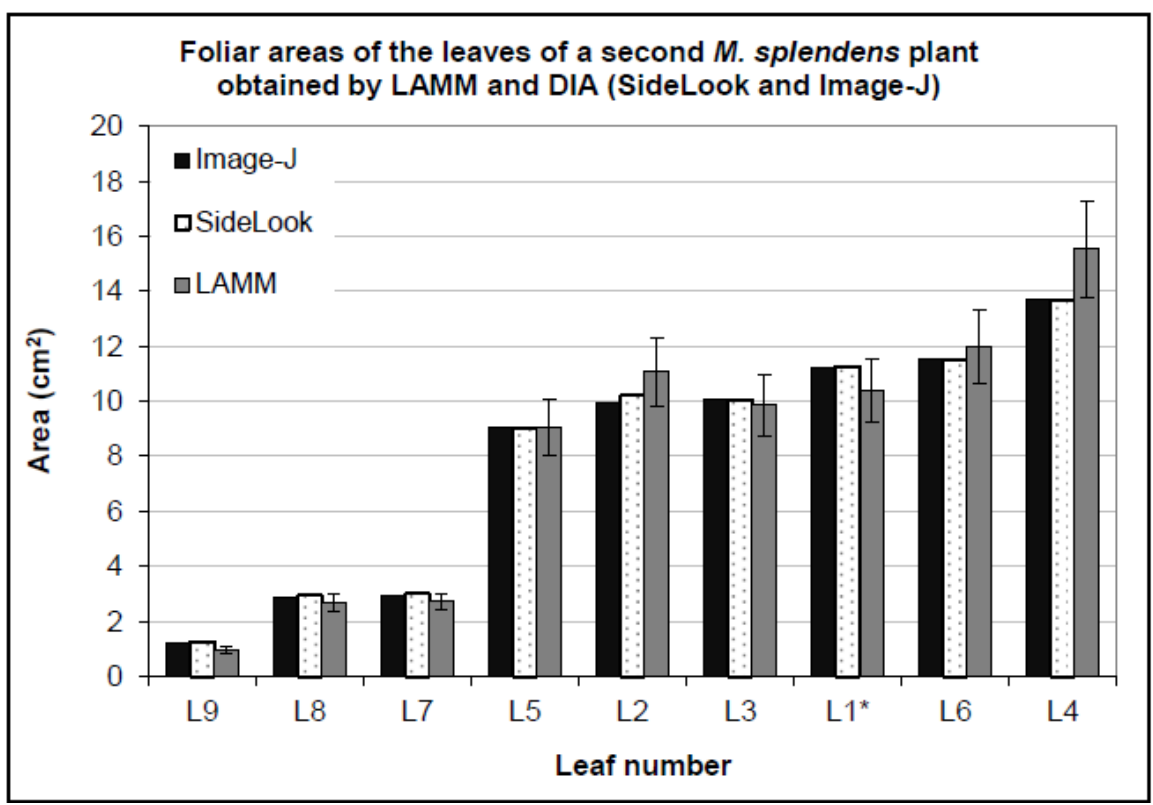

**Fig 5a** Foliar areas of 9 leaves of the plant *M. splendens*, L1 to L9, selected at random and measured by the techniques of DIA (IJ and SL) and by LAMM using the $K_G$ value obtained from a set of leaves of a different plant of the same species (from table 1). The L1 was a mature-chlorotic leaf (indicated as L1*). The average RD of the LAMM with respect to the SL measurements was $(8.50 \pm 6.90)\%$. The t-test between the LAMM and DIA values of the average area for the entire ensemble of leaves, showed no statistically significant difference.

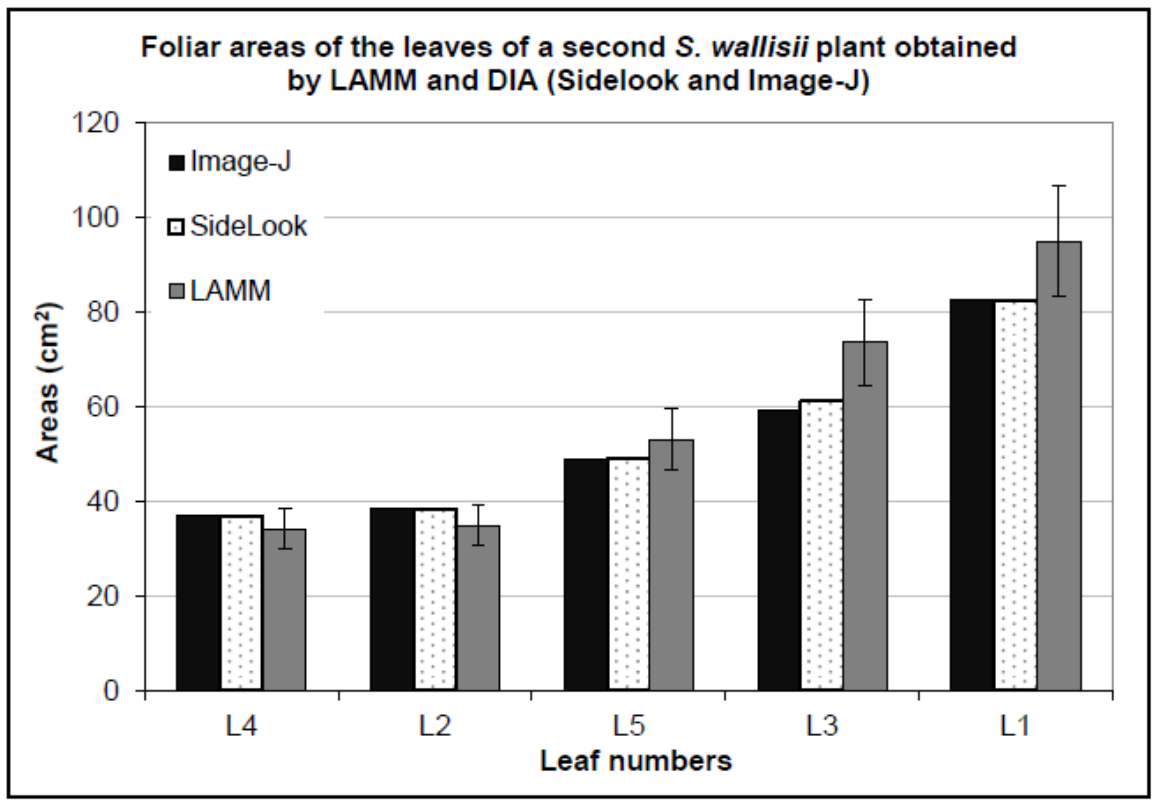

**Fig 5b** Foliar areas of 5 leaves of the plant *S. wallisii* selected at random and measured by the techniques of DIA (IJ and SL) and by LAMM using the $K_G$ value obtained from a set of leaves of a different plant of the same species (from table 1). The average RD of the LAMM with respect to the DIA measurements was $(12.08 \pm 5.43)\%$. The t-test between the LAMM and DIA values of the overall average area, showed no statistically significant difference

Figure 5a shows the LAMM and DIA measured areas of 9 leaves of a different *M. splendens* plant that also included one mature-chlorotic leaf. For the LAMM measurement, the $K_G$ value obtained from the *first* (and therefore different) *M. splendens* plant was used. Figure 5b shows the analogous measurements for the 5 leaves of a different *S. wallisii* plant where again the $K_G$ value from the *first S. wallisii* had been used. The t-test comparison between the overall average areas obtained by the LAMM and by the DIA in the case of both plants revealed that there was no statistically significant difference (95% CL) between the two types of measurement, just as in the case of the measurements in section 4.2 where the *intra*-plant $K_G$ value had been used. Further, the overall average RD between the LAMM and the DIA (Sidelook) were $(8.5 \pm 6.9)\%$ and $(12.1 \pm 5.4)\%$ for the *M. splendens* and *S. wallisii* respectively. A comparison between these RD values with those for the two plants of the same species of the first set (section 4.2) revealed that there was no statistically significant difference (95% CL) between them. Since the DIA measurements are independently performed on each plant (with no carry-over values such as the $K_G$), the similitude of the % average relative deviations for the two sets of the *M. splendens* and *S. wallisii* plants implies that the LAMM for the second set using the $K_G$ derived from the first set, produces results that are similar in the level of accuracy. This means that the $K_G$ value can be exported from one plant to another of the same species with little if any consequent loss in the degree of accuracy of the LAMM measurements. This exportability is the direct consequence of the conservation of the Hughes constant within a species (Roderick & Cochrane, 2002). The implication of this finding is that the LAMM method reduces to a very simple operation when the foliar area of a group of plants of the same species located in a stand is desired. The $K_G$ values can be extracted from a representative plant and exported to the others. In the interest of a homogeneous representation however, it would be preferable to obtain the $K_G$ value as the average of the Ks of leaves excised from various plants of the mono-specific set. For groups of mixed species, the process is replicated for each group.

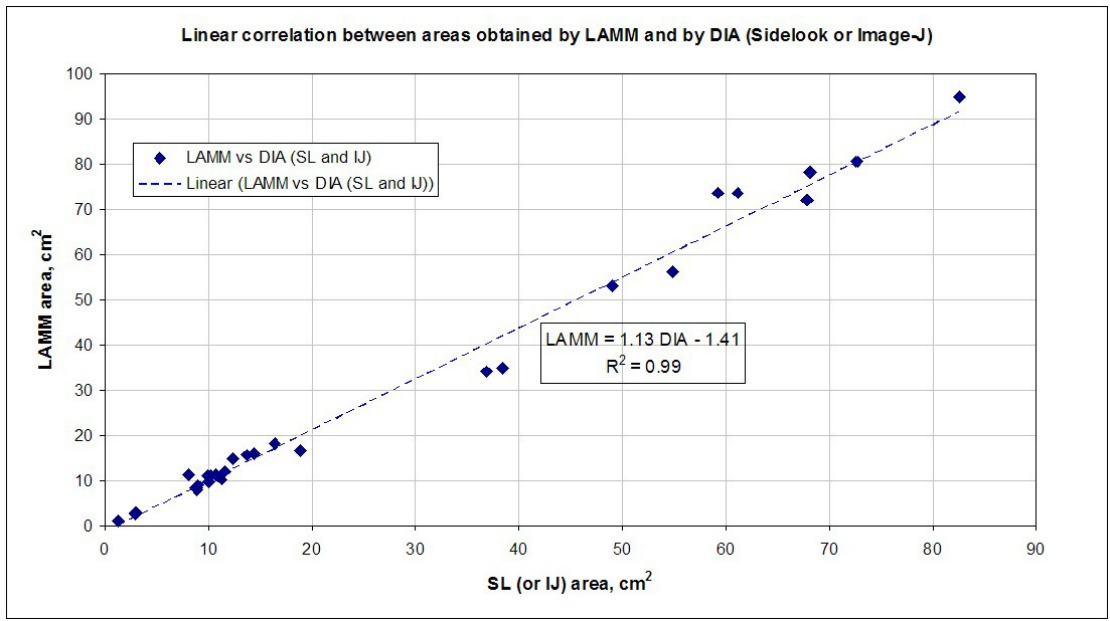

**Fig 6**   The linear correlation between the LAMM and the DIA measurements using Sidelook and Image-J of the leaf areas of the plants *M. splendens* and *S. wallisii*.

The degree of linear correlation between the DIA methods based on both SIDELOOK and IMAGE-J and the method of LAMM for all cases concerning leaves studied in this work, is shown in figure

6. As can be seen from the $R^2$ value (=0.99), the correlation is excellent.

*4.4     General considerations*

A number of previous workers (Sharatt & Baker, 1986; Ma *et al*., 1992; Vile *et al*, 2005) have utilized the concept of the SLA, leaf dry mass (LDM) and the dimensional attributes of leaves to obtain the quantification of a particular desired attribute. In the works of Sharatt & Baker (1986) and Ma *et al*. (1992), the leaf area (LA) is connected to the LDM via a regression fitted equation, while in the work of Vile *et al*. (2005) the leaf $\tau$ is connected to the inverse of the product of SLA and LDM by a regression equation whose coefficient they show is $\approx \rho^{-1}$. These methods come under the purview of the general gravimetric method discussed by Jonckheere et al (2004). However, they are all relative methods. They need a prior calibration via the exact measurement of the very attribute they seek to measure (LA or $\tau$) in the given plant species in order to obtain the fit coefficients. Strictly, from the standpoint of measurement methodology, such a cyclic argument is of limited utility. In contrast, the LAMM is an absolute method requiring no prior calibration fits. It measures the Hughes constant *in situ*. It can also be fine-tuned to any desired level of 'stratification'. The combination of the conservativeness of the Hughes constant and the idea of methodology borrowed from thick film technology permits the LAMM to be an absolute technique. By doing away with the calibration and drying steps in the gravimetric technique, LAMM substantially improves on the speed of operation. Thus LAMM may find utility in those cases where an alternative to the DIA without the disadvantages of the gravimetric method, is sought  (see section 1 and Jonckheere et al, 2004).

The Hughes' constant K is the pivot in the LAMM method. As is true for other biometric indices, K may change with different environmental conditions for the given species. Since the LAMM method measures the K value *in situ* for the given species, the variability of K under different environmental conditions becomes an irrelevant issue.

LAMM has been applied to other plants, such as *Aldama dentata* and *Tagetes erecta,* with and without mycorrhizal symbiosis, to determine the impact of soil contamination by copper and lead, and of the symbiosis with mycorrhiza, on the plant foliar areas (Dasgupta-Schubert et al, 2011; Castillo Baltazar 2011; Alvarado Lopez 2011). The LAMM measurements were selectively cross-checked by the DIA. In all experimental cases the LAMM foliar areas were found to follow the trend expected on the basis of the physiology of plants stressed by the aforesaid heavy metals and affected by the symbiosis with mycorrhizae. These studies serve to further validate the LAMM technique.

## 5. Conclusion and Perspectives

With the results of the three sets of experiments, the hypotheses stated in section 2 have been proved and the basis of LAMM established factually. The experiments also serve to corroborate in a quantitative manner the earlier observations of Roderick & Cochrane (2002). LAMM emerges as a simple yet novel technique for the measurement of the foliar area of broad-leaved herbaceous plants. It is also accurate, rapid and economic. The perspective for the future lies in the automation of its procedure with the use of high-quality die-punches of accurate IDs and with an ID range that extends to much lower sizes so that the areas of finely pinnated leaves may also be measured. The features might include not only the measurement of the FA but also the data acquisition and book-keeping of the Hughes constant whose statistically significant variation within a species, if detected, might arise from the failure of the regulatory process underlying its constancy, due to a variety of environmental stress factors. Furthermore, the method does not interfere with other optical methods of foliar analysis and could be added on as an additional feature to existing equipment.

While this work illustrates the application of LAMM to broad-leaved species, the principle underlying it is universal and applicable to the leaves of all plants, as long as the conservativeness of K for the species is not destroyed.

**Acknowledgements**: The authors are grateful to CONACyT of Mexico for funding their doctoral (OSC and CJA), master's (EMZ) and faculty (NDS) sabbatical-year (No. 173076) fellowships respectively. NDS thanks the Helmholtz Centre for Environmental Research, particularly Dr. Peter Kuschk, for the hospitality she received during her sabbatical leave there.